\documentclass[twocolumn,letter]{aa} 
\usepackage{graphicx}
\usepackage{txfonts}
\def\mo  {M$_\odot$}
\def\kms {\ifmmode{{\rm \ts km\ts s}^{-1}}\else{\ts km\ts s$^{-1}$}\fi}
\begin{document}
\title{A candidate protostellar object in the L\,1457 / MBM\,12 cloud}
\authorrunning{Heithausen \& B\"ottner}
\titlerunning{A candidate protostellar object
 in the  L1457 / MBM12 cloud }
\author{Andreas Heithausen\inst{1}\and
        Christoph B\"ottner\inst{2}}
\offprints{A. Heithausen}
\institute{ 
Institut f\"ur integrierte Naturwissenschaften, Abteilung Physik,
Universit\"at Koblenz-Landau, Universit\"atsstr. 1, 56070 Koblenz,
Germany,
 \email{heithausen@uni-koblenz.de},
\and
Radioastronomisches Institut, Universit\"at Bonn,
Auf dem H\"ugel 71, 53121 Bonn, Germany,
 \email{christoph-boettner@gmx.de} 
   }
\date{Received 3 March 2010; accepted 30 September 2010 }

\abstract
{}
{The association of young T\,Tauri stars, MBM\,12A, indicates that
L\,1457 was forming stars not too long ago. With our study we want to
find out whether there are still signs of ongoing star formation in
that cloud.}  
{Using the Max-Planck-Millimeter-Bolometer MAMBO at the
IRAM 30m telescope, we obtained a map of about $8'\times 8'$ centered
on L\,1457 in the dust continuum emission at 230 GHz. 
 Towards the most intense regions in our bolometer map, we
obtained spectra at high angular resolution in the CS $(2\to1)$ and
the N$_2$H$^+$ $(1\to0)$ lines using the IRAM 30m telescope.  }
{We find that the cold dust in L\,1457 is concentrated in several
small cores with high H$_2$ column densities and solar masses. The
density profiles of the cores are inconsistent with a sphere with
constant density.  These cores are closer to virial equilibrium than is
the cloud as a whole. Data from the VLA and Spitzer archives reveal
two point sources in the direction of one dust core. One of the
sources is probably a distant quasar, whereas the other source is 
projected right on a local maximum of our dust map  and shows
characteristics of a protostellar object. }
{}
   
\maketitle


\section{Introduction}

Class 0 objects are protostellar objects at the beginning of the
accretion phase, where the bulk of the final stellar material has not
been assembled yet (Andr\'e et al. \cite{andre:etal93}). These
objects have been detected towards various regions with known ongoing
star formation (Andr\'e et al. \cite{andre:etal93},
\cite{andre:etal99}; Kauffmann et al. \cite{kauffmann:etal05}).
Molecular cirrus clouds at high Galactic latitudes  so far show only
very little star-formation activity, no embedded infrared sources had
been found in the IRAS faint source catalog (Magnani et
al. \cite{magnani:etal95}). Nevertheless, the existence of 
a gravitationally bound core in a cirrus clouds 
(Heithausen et al. \cite{heithausen:etal02}, \cite{heithausen:etal08}),
which shows evidence of inward motion (Heithausen
\cite{heithausen99}), shows that these clouds harbor potential sites of
star formation.

\begin{figure*}
\sidecaption
\includegraphics[angle=0,width=11cm]{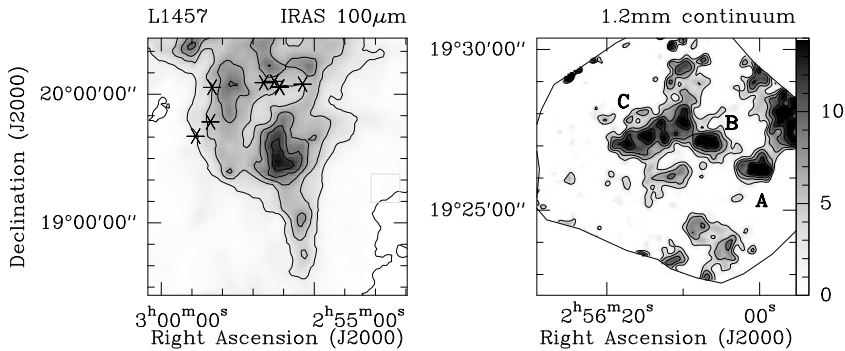}
\caption
{IRAS 100$\mu$m and MAMBO 1.2mm maps of L\,1457. In the IRAS map 
contour lines are every 5 MJy/sr starting at 5
MJy/sr.  The position of our CS and bolometer maps is marked by the
white box. The asterisks mark the positions of the T Tauri stars in
the L\,1457/MBM\,12 region described by Luhman (\cite{luhman01}).
 Contours in the MAMBO 1.2mm dust emission map are every 3
mJy/beam starting at 3 mJy/beam $(3\sigma)$. The positions of the
three condensations in the dust map are labeled A, B, and C. }
\label{l1457-dust}
\end{figure*}

One candidate for possible star formation in such clouds is the dark cloud
L\,1457, located at southern Galactic latitudes below the Taurus
region. This cloud is puzzling for two reasons:
\begin{itemize}
\item Complete CO maps of L\,1457 (Magnani et al.
\cite{magnani:etal85}; Zimmermann \& Ungerechts
\cite{zimmermann:etal90}; Pound et al. \cite{pound:etal90}) indicate
that the cloud as a whole is far from being gravitationally bound.
Reach et al. (\cite{reach:etal95}) found extensive CS emission
throughout the cloud, indicating dense molecular gas that is
closer to virial equilibrium. 

\item Several T\,Tauri stars have been detected in the
direction of L\,1457 (Hearty et al. \cite{hearty:etal00}; Luhman
\cite{luhman01}), forming the association MBM\,12A (see 
Fig. \ref{l1457-dust}).  For this assocation, Luhman (\cite{luhman01})
determined an age of about 2 million years, indicating that, if the
young stars are indeed associated with the cloud, L\,1457 was forming
stars not too long ago.
\end{itemize}

To shed light on its star-forming capability, we obtained a bolometer
map of the densest part of the L\,1457 / MBM\,12 cloud at a wavelength of
1.2\,mm with the IRAM 30\,m telescope. The selected region was found to be the
most intense portion  in the high angular resolution data of the $^{12}$CO and
$^{13}$CO $(1\to 0)$, $(2\to 1)$, and $(3\to 2)$ transitions 
obtained by Zimmermann
(\cite{zimmermann93}).  Supplementary information on the dynamical state
of the core comes from CS $(2\to1)$ observations obtained with the 
IRAM 30m telescope, too. We discuss a
possible association of two continuum point sources detected with the
VLA at a wavelength of 3.6cm and with Spitzer Space Telescope at
24$\mu$m.

\section{Observations\label{observations}}

L\,1457 was observed in 2000 and 2001 in the dust continuum at 1.2mm
using the Max-Planck-Millimeter-Bolometer MAMBO at the IRAM 30m
radio telescope on Pico Veleta, Spain.  MAMBO is sensitive to emission
between 210 and 290\,GHz, with an effective frequency of 250\,GHz for
steep thermal spectra. The observations were taken in double-beam
on-the-fly mode, i.e., chopping the secondary mirror in azimuth by
50$''$ to 70$''$ at 2Hz and scanning the sky in azimuth at a speed of
4$''$ to 5$'' s^{-1}$, then moving in elevation by $4''$. The maps
were taken under variable winter conditions with line-of-sight
opacities between 0.2 and 0.7. The effective beam FWHM is 11$''$.

The most intense dust positions in L\,1457 were oberved in the CS
$(2\to1)$ transition in October 2002 and in the N$_2$H$^+ (1\to 0)$
transition in August 2004 with the IRAM 30m radiotelescope. Data were
taken in single-position on-off-mode. We obtained small maps with
20$''$ spacing between individual positions.  We used the VESPA
autocorrelation spectrometer with a velocity resolution set to 0.062
km s$^{-1}$ at 93GHz and 0.060 km s$^{-1}$ at 98GHz.  The angular
resolution of the 30m telescope at 93GHz and 98 GHz is 27$''$ and 
26$''$, respectively. The main beam efficiency $\eta_{mb}=0.8$.

\begin{figure*}
\sidecaption
\includegraphics[angle=0,width=11cm]{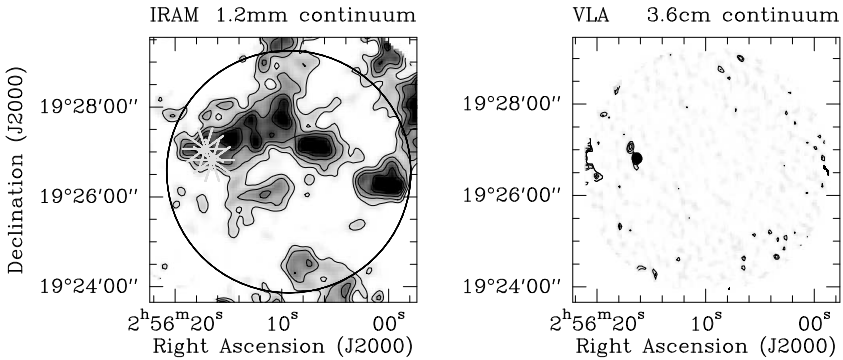}
\caption
{Comparison of the VLA map (right) at $\lambda=3.55$\,cm and our
bolometer map (left).  The circle in the bolometer map indicates the
primary beam of the VLA map. The positions of the VLA sources are
marked by asterisks. Contours in the bolometer map are the same as in
Fig. \ref{l1457-dust}. Contours in the VLA map are every 0.07
mJy/beam starting at 0.07 mJy/beam (2 sigma).}
\label{vla}
\end{figure*}

Additionally, we used public data from the VLA and Spitzer archives.
In the VLA archive we found two thus far unpublished maps centered on
about the same position as our bolometer map. The data were observed
on March 6 and 25, 1991 at a wavelength of 3.55\,cm with an angular
resolution of $10''$.  The primary beam of the VLA has a full width at
half maximum of $5.\!'4$ and covers all our bolometer sources
(s. Fig. \ref{vla}).  The noise level in the VLA map is 0.03
mJy/beam.  Part of the region was also observed with MIPS (Rieke et
al. \cite{rieke:etal04}) onboard the Spitzer Space Telescope
(Werner et al. \cite{werner:etal04}) at a wavelength of 24\,$\mu$m.
The data were observed on September 23, 2007.  The angular
resolution of the MIPS instrument at that wavelength is about 6$''$,
map spacing is 2.5$''$. The data taken from the Spitzer archive were
already calibrated using the standard pipeline. The noise level
in the MIPS map is 0.06 MJy/sr.

\begin{table*}
\caption{1.2mm sources in L\,1457}
\begin{tabular}{l l l l l l l l l l }
\noalign{\hrule}
\noalign{\smallskip}
 Source & RA       & Dec   & size & Radius & Peak $F_\nu^{beam}$ & Peak $N_{\rm H_2}$ & $F_\nu$ & $M_{\rm H_2}$ &
Virial Mass  \\
           & (J2000)  & (J2000)  &($''\times''$) & (pc) &(mJy/beam)& (cm$^{-2}$) & (Jy)  & (\mo) & (\mo)  \\
\noalign{\smallskip}
\noalign{\hrule}
\noalign{\smallskip}
L\,1457-A   & 2:56:00.6 & 19:26:17 &  $67\times46$ & 0.07 & $20\pm1$ & $1.3 \times 10^{22}$& $0.23\pm0.03$ & 1.0 & $\le 3  $  \\
L\,1457-B   & 2:56:07.0 & 19:27:10 &  $57\times40$ & 0.06 & $17\pm1$ & $1.1 \times 10^{22}$& $0.17\pm0.02$ & 0.7 &  $\le 4 -10$\\
L\,1457-C   & 2:56:13.5 & 19:27:30 & $146\times56$ & 0.11 & $17\pm1$ & $1.1 \times 10^{22}$& $0.74\pm0.05$ & 3.1 & $\le 5 $   \\
L\,1457-C1   & 2:56:17.2 & 19:27:04 & $54\times55$ & 0.08 & $17\pm1$ & $1.1 \times 10^{22}$& $0.21\pm0.03$ & 0.9 & $\le 3$ \\
\noalign{\smallskip}
\noalign{\hrule}
\end{tabular}
\label{sourcesl1457}
\end{table*}

\begin{figure*}
\includegraphics[angle=-90,width=18cm]{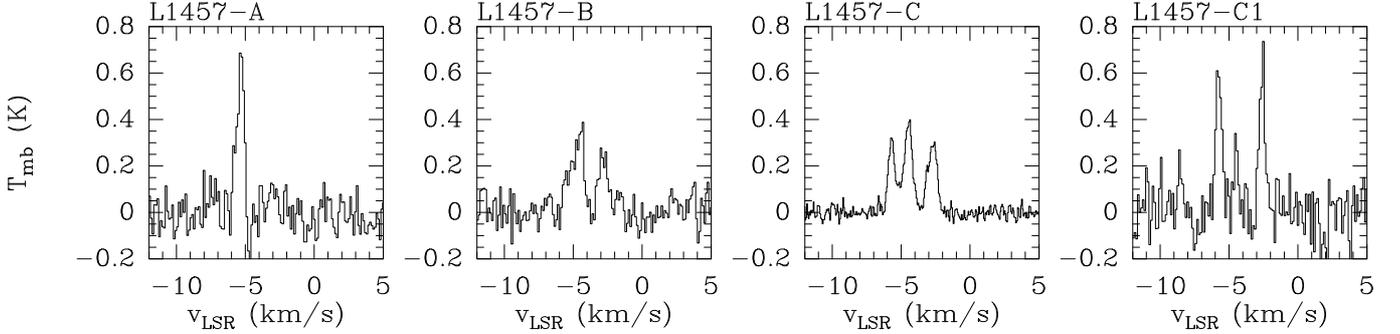}
\caption
{CS spectra towards the L\,1457-A, L\,1457-B, and L\,1457-C,  as
obtained with the IRAM 30m telescope. For L\,1457-A and B, we only obtained
single spectra towards the center of the cores. For L\,1457-C we
obtained a large map with 50 individual positions, which we
averaged. The rightmost spectrum shows a single CS spectrum towards
the position of J025616+192703.}
\label{l1457-abc}
\end{figure*}

\section{The distribution of cold dust\label{results}}

Our MAMBO 1.2mm dust map is presented in Fig. \ref{l1457-dust}.  Dust
is concentrated in several small condensations. A possible
extensive diffuse dust component is filtered out by our observing
technique. We labeled the three most prominent structures as L\,1457-A,
B, and C. The part of the core where we detected a point source
(see Sect. \ref{sources}) is referred to as L1457-C1.

To calculate the molecular hydrogen column density ${N_{\rm H_2}}$
from the peak flux density $F_\nu^{beam}$\ we use the formula provided
by Kauffmann et al. (\cite{kauffmann:etal08}), who adopted a dust
temperature of 10\,K and a wavelength of 1200\,$\mu$m:

${N_{\rm H_2}} = 6.69 \cdot 10^{20} {\rm cm^{-2}}
\left(\frac{\displaystyle F_\nu^{beam}} {\displaystyle \rm mJy\
beam^{-1}}\right)$.

\noindent Ammonia observations reveal a gas temperature of about $T_{kin}\le
12\, \rm K$ for the dense core (Gomez et al. \cite{gomez:etal00}).  We
derive molecular hydrogen column density of up to $N_{\rm H_2}=1.3
\times 10^{22}$\, cm$^{-2}$ (s. Table \ref{sourcesl1457}).  Towards
XY\,Ari, a star in the middle of our field, Littlefair et
al. (\cite{littlefair:etal01}) derived a visual extinction of
$A_V=11.5\pm0.3$\,mag, confirming our estimate of the molecular
hydrogen column density.

Kauffmann et al.  (\cite{kauffmann:etal08}) also provide 
a formula for deriving the molecular
hydrogen mass ${M_{\rm H_2}}$\ from the total bolometer flux $F_\nu$
at a wavelength of 1200$\mu$m for an object with 10\,K temperature
which is applicable to our observations

${M_{\rm H_2}} = 0.47\ {\rm M_\odot} \left(\frac{\displaystyle
F_\nu}{\displaystyle\rm Jy\ }\right) \left(\frac{\displaystyle
d}{\displaystyle 100 \rm pc}\right)^2$

\noindent where $d$ is the distance to the cloud.  
With a distance of $d=65$\,pc,
L\,1457 has long been thought to be the nearest molecular cloud (Hobbs
et al. \cite{hobbs:etal86}). Recent distance determinations suggest,
however, a much higher distance.  Andersson et
al. (\cite{andersson:etal02}) estimate a distance of $360\pm30$pc to
L\,1457; Luhman (\cite{luhman01}) finds 275\,pc, and Strai\v{z}ys et
al.  (\cite{straizys:etal02}) find 325\,pc. Adopting a distance of 300\,pc 
to the cloud, we find masses in the range of $0.7\,{\rm M_\odot}\le
M_{\rm H_2}\le3.1\,\rm{M_\odot}$ (Table \ref{sourcesl1457}).

Similar to our approach for MCLD123.5+24.9 (Heithausen et
al. \cite{heithausen:etal08}) we tried to determine the density
profiles of L1457-A and B by obtaining circular averages of the
intensity distributions centered on the cores. We find that both are
inconsistent with spheres of constant volume densities. Better fits
can be derived by Gaussian profiles or centrally peaked density
profiles similar to those given by Johnstone et
al. (\cite{johnstone:etal03}, see also Heithausen et
al. \cite{heithausen:etal08}).

To analyze the stability of the cores we compare their masses with
 their virial masses 
$M_{\rm vir} = 3 r \sigma^2 / G$ , where $\sigma = \Delta v
/ 2.355$ is the one-dimensional gas velocity dispersion, $r$\ the
radius of the cloud, and $G$ the gravitational constant. 
Here we assume a density profile proportional to $r^{-2}$. 
Because towards the regions with significant dust emission no
N$_2$H$^+$ was detected down to a rms of $\le0.06$\,K $(T_{mb})$ in
0.06\,\kms\ wide channels, we use CS $(2\to1)$ data obtained at high
angular resolution with the IRAM 30\,m telescope
(s. Fig. \ref{l1457-abc}) to estimate the velocity dispersion.
Parameters of a Gaussian analysis of the CS spectra are listed in Table
\ref{l1457-cs-abc}. For L\,1457-B and C, the CS spectra show multiple
components.  For our estimate of the virial mass we took the
components closest to $v_{\rm LSR}=-5$\,\kms, which is the velocity of
the main CS component.  If we correct for the contribution of Helium,
the ratio of cloud mass to virial mass is in the range of 0.2 to 0.6,
indicating that the cores are close to virial equilibrium. This ratio
is a lower limit because CS traces a larger volume in the core (e.g. Reach
et al. \cite{reach:etal95}) than our bolometer sources 
Furthermore the CS line might be optically thick.
We might therefore systematically overestimate the line
width, hence also the virial mass.

\begin{table}
\caption{Parameters of the IRAM CS observation in L\,1457}
\begin{tabular}{l l l l l }
\noalign{\hrule}
\noalign{\smallskip}
 Source & $T_{mb}$ & rms & $v_{LSR}$ & $\Delta v$ \\
        & (K)     & (K) & (\kms)    & (\kms) \\
\noalign{\smallskip}
\noalign{\hrule}
\noalign{\smallskip}
L\,1457-A   & 0.69 & 0.09 & $-5.35\pm0.02$ & $0.56\pm0.04$\\
L\,1457-B   & 0.31 & 0.08 & $-4.71\pm0.05$ & $1.17\pm0.11$\\
          & 0.25 & 0.08 & $-2.82\pm0.05$ & $0.74\pm0.10$\\
L\,1457-C   & 0.37 & 0.02 & $-5.69\pm0.01$ & $0.61\pm0.03$ \\
          & 0.31 & 0.02 & $-4.49\pm0.01$ & $0.74\pm0.03$\\
          & 0.30 & 0.02 & $-2.74\pm0.01$ & $0.79\pm0.03$\\
L\,1457-C1   & 0.62 & 0.10 & $-5.75\pm0.03$ & $0.50\pm0.06$ \\
          & 0.33 & 0.10 & $-4.49\pm0.04$ & $0.31\pm0.11$\\
          & 0.64 & 0.10 & $-2.59\pm0.03$ & $0.48\pm0.06$\\
\noalign{\smallskip}
\noalign{\hrule}
\end{tabular}
\label{l1457-cs-abc}
\end{table}

\section{Radio and infrared sources towards the core of 
L\,1457 / MBM\,12 \label{sources}}
The VLA 3.6cm maps (Fig. \ref{vla}) show two point
sources. The brighter one has already been detected by Gomez et
al. (\cite{gomez:etal00}) from other VLA archive data and they named it
VLA\,B0253+192. About $18''$ to the northeast we detected a
further point source, which we name J025616+192703.  Positions and
fluxes are listed in Table \ref{l1457-sources}.

For VLA\,B0253+192, Gomez et al.  (\cite{gomez:etal00}) give a flux
density of $S_{\rm 8.4GHz}=0.114 \pm 0.026$\,mJy at the observed
wavelength. We determine a higher flux $S_{\rm 8.4GHz}=0.60 \pm
0.04$\,mJy, so the source might be variable.  The source is also
listed in the NVSS catalog by Condon et al.  (\cite{condon:etal98}) at
a frequency of 1.4 GHz.  At that frequency Condon et al. give a flux
density of $S_{\rm 1.4GHz}=3.4 \rm mJy$.  Using the flux density by
Gomez et al.  (\cite{gomez:etal00}) we find a spectral index of
$\alpha_{\rm Radio}=-1.9$, while  we find $\alpha_{\rm
Radio}=-1.0$ using our value.
 Both values indicate a steep radio spectrum typical for
non-thermal emission. This confirms the proposal by Gomez et al. that
the source might be an unrelated background source, possibly a distant
quasar.

For J025616+192703 no counterpart can be found in the NVSS
catalog.  Due to the angular resolution of NVSS catalog of 45$''$
(Condon et al. \cite{condon:etal98}), this source is at only half a
beam distance from VLA B0253+192. To separate a possible contribution
of J025616+192703, we subtracted the emission of VLA B0253+192
from the NVSS image adopting a circular beam size. This way we find an
upper limit of 0.3 mJy for the flux of J025616+192703 at 1420\,MHz.
The spectral index derived with that value would close to or larger than zero, 
thus
synchrotron radiation can be excluded. The spectrum is thus clearly
different from that of VLA\,B0253+192, making a physical association
between both sources unlikely.

\begin{figure}
\includegraphics[angle=0,width=9cm]{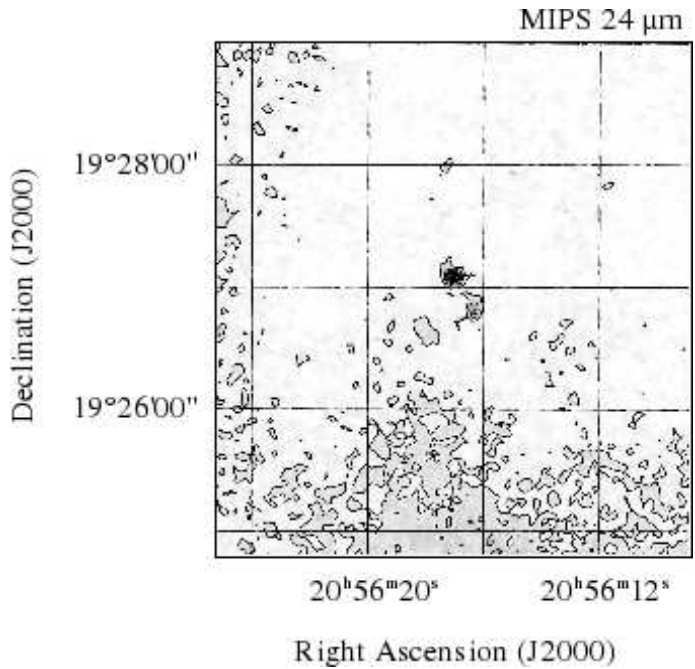}
\caption
{24$\mu$m Spitzer map of the dense core in L\,1457. Contours are every
0.2 MJy/sr. Note that the declination and right ascension axes are
 slightly rotated, as indicated by the grid.}
\label{l1457-mips}
\end{figure}

\begin{table}
\caption{Parameters of the VLA and MIPS sources in L\,1457}
\begin{tabular}{l l l}
\noalign{\hrule}
\noalign{\smallskip}
        & J025616+192703 & VLA\,B0253+192  \\
\noalign{\smallskip}
\noalign{\hrule}
VLA 21\,cm$^1$ \\
\noalign{\hrule}
$\alpha(\rm J2000)$ & $ - $ &  $2^{\rm h}56^{\rm m}16^{\rm s}.0$  \\
 $\delta(\rm J2000)$ & $-$& $ 19^\circ26'50''\!.8$\\
 $S_{\rm 1.4GHz}$ & $\le0.3$\,mJy& $3.4\pm0.5$\,mJy\\
\noalign{\smallskip}
\noalign{\hrule}
VLA 3.6\,cm\\
\noalign{\hrule}
$\alpha(\rm J2000)$ & $ 2^{\rm h}56^{\rm m}16^{\rm s}.9$ &  $2^{\rm h}56^{\rm m}16^{\rm s}.3$  \\
 $\delta(\rm J2000)$ & $19^\circ27'03''\!.6$& $ 19^\circ26'48''\!.8$\\
 $S_{\rm 8.4GHz}$ & $0.24 \pm 0.04$\,mJy& $0.60\pm0.04$\,mJy\\
\noalign{\smallskip}
\noalign{\hrule}
MIPS 24\,$\mu$m\\
\noalign{\hrule}
$\alpha(\rm J2000)$ & $ 2^{\rm h}56^{\rm m}17^{\rm s}.0$&  $2^{\rm h}56^{\rm m}16^{\rm s}.3$  \\
 $\delta(\rm J2000)$ & $19^\circ27'03''\!.9$&  $ 19^\circ26'48''\!.6$\\
 $S_{\rm 24\mu m}$ & $2.0 \pm 0.2$\,mJy& $1.0 \pm 0.2$\,mJy \\
\noalign{\smallskip}
\noalign{\hrule}
\noalign{\smallskip}
\noalign{Remarks: 1: Condon et al. (\cite{condon:etal98})}
\end{tabular}
\label{l1457-sources}
\end{table}

Both sources are also visible in the Spitzer 24\,$\mu$m map (see
Fig. \ref{l1457-mips}). Fluxes and positions are given in
Table \ref{l1457-sources}. The Spitzer maps at 70\,$\mu$m and
170\,$\mu$m are unfortunately disturbed and cannot be used to
determine the infrared spectra of the sources. In the IRAS 100\,$\mu$m
map and our bolometer map the emission is extensive, so we can only
determine upper limits for the source fluxes at those wavelengths.

As can be seen in Fig. \ref{vla} the source J025616+192703 is
projected precisely on a local emission maximum in our bolometer map
of L\,1457-C. A chance alignment of a background source is
very low. It is therefore possible that this source is indeed
associated with the dense core, possibly a protostellar condensation,
as a Class 0 object, deeply embedded in the core. 

The few points of the spectrum of J025616+192703 that we know so far are
consistent with established Class 0 sources.  For example,
IRAM\,04191+1522 (Andr\'e et al. \cite{andre:etal99}) would have a
similar infrared intensity at 24\,$\mu$m (Dunham et
al. \cite{dunham:etal06}), if put at the same distance. This source is
also seen at radio wavelength with a similar flux (Andr\'e et
al. \cite{andre:etal99}).  Recently, Kauffmann et
al. (\cite{kauffmann:etal05}) and Bourke et al.
(\cite{bourke:etal06}) have detected protostellar objects with very low
luminosities. These sources cannot be distinguished from the extensive
dust emission at mm-wavelength, but are more easily detected at shorter 
wavelengths, similar to J025616+192703. Without further
points in the spectrum it is, however, not possible to draw more
conclusions about the evolutionary state about our source.

\section{Summary and conclusions}

We have presented a bolometer map at 1.2mm, which shows at least three
dense dust condensations with peak H$_2$ column densities of
$\ge10^{22}$\,cm$^{-2}$ and solar masses. These are closer to virial
equilibrium than is the source as a whole. 

Towards one of the condensations, we find two point sources seen in the
radio regime at 3.6\,cm and in the infrared at 24\,$\mu$m. The radio
spectral index of one of the sources indicates nonthermal emission,
thus the source is most likely a background object.  For the other
source the spectrum is only sparsely known. The exact location of the
source at a local maximum in L\,1457-C suggests a physical assocation
with the cloud. We speculate that this source could be a protostellar
condensation that is still deeply embedded in the core. Its low luminosity and
the detection only at radio and infrared wavelengths could be caused
by an object with a temperature similar to that of the surrounding
material.

To support this hypothesis more observations of the spectrum of this
object between radio and infrared wavelength or a search for an
associated outflow are necessary.  Possibly owing to the multiple
velocity components in the CO data of L1457 
(Zimmermann \cite{zimmermann93}) no such
outflow has been detected so far. If the
protostellar nature of the source is confirmed, the study of this source
 will significantly
increase our knowledge of low-mass star formation outside of the Galactic plane.

\begin{acknowledgements}
This work is based on observations carried out with the IRAM 30\,m
telescope and the VLA. IRAM is supported by INSU/CNRS (France), MPG
(Germany), and IGN (Spain). The National Radio Astronomy Observatory is
a facility of the National Science Foundation operated under
cooperative agreement by Associated Universities, Inc. This work is
also based in part on archival data obtained with the Spitzer Space
Telescope, which is operated by the Jet Propulsion Laboratory,
California Institute of Technology under a contract with NASA.
\end{acknowledgements}

\end{document}